\documentclass[final,5p,times]{elsarticle}
\usepackage{epsfig}
\usepackage{latexsym}

\usepackage[colorlinks=true,linktocpage=true,linkcolor=blue,citecolor=blue,allcolors=blue]{hyperref}
\usepackage{url}
\usepackage[utf8]{inputenc}

\usepackage{enumerate}
\usepackage{color}
\usepackage{xcolor}

\usepackage{xfrac}

\usepackage{amsmath}
\usepackage{amssymb}

\usepackage[english]{babel}
\usepackage{hyperref}
\usepackage{url}
\usepackage{needspace}
\usepackage{float}
\hypersetup{
   colorlinks=true, 
  linktoc=all,     
  linkcolor=blue,  
}

\DeclareMathOperator{\EX}{\mathbb{E}}

\begin{document}

\begin{frontmatter}

\title{Non-monotonicity of $p_T$ correlations from meson-baryon mixing}

\author[fifth]{Tom Reichert}

\author[first,second]{Jan Steinheimer \corref{cor1}}
\ead{j.steinheimer-froschauer@gsi.de}
\cortext[cor1]{Corresponding author}
\affiliation[first]{organization={GSI Helmholtzzentrum f\"ur Schwerionenforschung GmbH},
            addressline={Planckstr. 1}, 
            postcode={D-64291}, 
            city={Darmstadt},
            country={Germany}
}

\affiliation[second]{organization={Frankfurt Institute for Advanced Studies (FIAS), Ruth-Moufang-Str. 1, D-60438 Frankfurt am Main, Germany}}

\author[third,first,fourth]{Marcus Bleicher}

\affiliation[third]{organization={Institut für Theoretische Physik, Goethe-Universit\"{a}t Frankfurt, Max-von-Laue-Str. 1, D-60438 Frankfurt am Main, Germany}}
\affiliation[fourth]{organization={Helmholtz Research Academy Hesse for FAIR (HFHF), GSI Helmholtzzentrum f\"ur Schwerionenforschung GmbH, Campus Frankfurt, Max-von-Laue-Str. 12, 60438 Frankfurt am Main, Germany}}

\affiliation[fifth]{organization={Theoretical Physics Department, CERN, 1211 Geneva 23, Switzerland}}

\begin{abstract}
The STAR experiment has recently reported data on the $\sqrt{\langle\Delta p_{T,i}\Delta p_{T,j}\rangle}/\langle\langle p_T\rangle\rangle$ charged hadron correlation in Au+Au reactions from $\sqrt{s_{NN}}=3-200$ GeV. The beam energy dependence of this quantity is non-monotonic, showing a pronounced minimum at $\sqrt{s_{NN}} \approx 7.7$ GeV, while being essentially flat at lower and higher energies. It has been proposed that such a non-monotonicity would be consistent with increased momentum correlations due to a critical point of QCD. In the present work it is shown, using a simplified model, that the observed structure can be consistently explained by the transition from a baryon dominated system to a meson dominated system and is therefore not a good observable for the critical point of QCD.
\end{abstract}


\end{frontmatter}


\section{Introduction}
Quantum-Chromo-Dynamics (QCD) is the fundamental theory describing strongly interacting particles. I.e. QCD governs the interactions and properties of matter on nuclear and subatomic scales. On the theoretical side, ab-initio calculations based on lattice QCD allows to explore the properties of strongly interacting matter along the temperature ($T$) axis of the phase diagram at vanishing and moderate baryo-chemical potentials $\mu_B$. These calculations have shown that one expects a crossover between confined and deconfined matter in the explored $\mu_B/T<3$ region of the phase diagram \cite{Borsanyi:2010bp,Bazavov:2011nk,Bazavov:2017dus,Borsanyi:2020fev,HotQCD:2018pds,Vovchenko:2017gkg}. At the moment it is speculated that this crossover may end in a second order phase transition at a specific $T_{\rm CEP},\mu_{\rm B, CEP}$ denoting a critical end point (CEP). Studies based on Dyson–Schwinger equations and functional renormalization group methods suggest the CEP to be located at $T^{\rm CEP}=80-140$ MeV, $\mu_B^{\rm CEP}=500-800$ MeV \cite{Fischer:2014ata,Fu:2019hdw,Gao:2020qsj,Gunkel:2021oya}, which can be roughly matched to a center-of-mass collision energy of $\sqrt {s_{NN}}=4-8$ GeV.

On the experimental side, this energy range has been extensively explored by the RHIC-BES program, which is currently analyzing the data taken for Au+Au reactions during the previous run times. A central goal of this analysis is to find signature of the CEP. Typically, one assumes that at and near the critical region the correlation length of the system diverges and fluctuations and correlations increase drastically \cite{Stephanov:1998dy,Hatta:2003wn}. While this is evident for macroscopic systems that can be held at at fixed thermodynamic parameters for an arbitrary duration, it is not clear that a large signal will emerge in a small transient system like a heavy ion collision \cite{Paech:2003fe,Nahrgang:2011vn,Herold:2013bi}. Another complication arises from the fact that heavy ion experiments can only measure correlations in momentum space while the correlations due to a phase transition and a critical endpoint appear firstly in coordinate space.

The STAR experiment has recently reported data on the normalized dynamical $p_T$ fluctuation of the transverse momentum of charged particles $\sqrt{\left<\Delta p_{T,i}\Delta p_{T,j}\right>}/\left<\left< p_T\right>_k\right>$  in central and peripheral Au+Au reactions from ${\sqrt{s_{NN}}=3-200}$GeV \cite{STAR:2019dow,STAR:2026vjv}. Fluctuations of the transverse momentum have been discussed in the literature in various works \cite{Mrowczynski:2004cg,Korus:2001au,Mrowczynski:2004cg,NA49:2003hxt,ALICE:2014gvd,BRONIOWSKI2006290,Jeon:2003gk,Gavin:2016nir,Cody:2021cja,Zhang:2025yyd,Du:2025tdh}.
STAR reports that the energy and centrality dependence of this observable \cite{STAR:2005vxr,STAR:2026vjv} is non-monotonic, showing a pronounced minimum around $\sqrt{s_{NN}}=4-10$GeV, while being essentially flat at lower and higher energies. They speculate that this non-monotonicity could be a sign of the critical point of QCD. The aim of this letter is to provide a clearer interpretation of this dynamical $p_T$ fluctuation and to elucidate in a very simplified approach how such a structure can emerge by a change from a baryon dominated system to a meson dominated system.
Thus, questioning the interpretation of the observed minimum as a signature of a CEP.

\section{Physical interpretation of the correlator \texorpdfstring{$\left<\Delta p_{\rm T,i}\Delta p_{\rm T,j}\right>$}{<Delta pT,i Delta pT,j>}}

To characterize the $p_{\rm T}$ correlations, STAR \cite{STAR:2019dow,STAR:2026vjv} used the two-particle $p_{\rm T}$ covariance given by:
\begin{equation}
	\left<\Delta p_{{\rm T},i}, \Delta p_{{\rm T},j}\right> = \frac{1}{N_{\rm ev}}\sum_{k=1}^{N_{\rm ev}}\frac{C_k}{N_k\left(N_k - 1\right)},
	\label{eq:dptdpt}
\end{equation}
where
\begin{equation}
	C_k = \mathop{\sum_{i,j=1}^{N_k}}_{i\neq j}\left(p^{(k)}_{{\rm T},i} - \left<\left<p_{\rm T}\right>_k\right>\right)\left(p^{(k)}_{{\rm T},j} - \left<\left<p_{\rm T}\right>_k\right>\right).
	\label{eq:Ck}
\end{equation}
Here $N_{\rm ev}$ is the number of events, $N_k$ is the number of charged particles in event $k$, and $p^{(k)}_{{\rm T},i}$ is the transverse momentum of the charged particle $i$ in the given event $k$. The event-averaged mean transverse momentum $\left< p_{\rm T}\right>_k$ is defined as 
\begin{equation}
	\left<\left<p_{\rm T}\right>_k\right> = \frac{\sum_{k=1}^{N_{\rm ev}}\left<p_{\rm T}\right>_k}{N_{\rm ev}},
	\label{eq:mmpt}
\end{equation}
where $\left<p_{\rm T}\right>_k$ is the average $p_{\rm T}$ of event $k$ defined as
\begin{equation}
	\left<p_{\rm T}\right>_k = \frac{\sum_{i=1}^{N_k}p^{(k)}_{{\rm T},i}}{N_k}.
	\label{eq:mpt}
\end{equation}
Similar analysis were performed by CERES \cite{CERES:2003sap} and ALICE \cite{ALICE:2014gvd}. 

To obtain a better understanding of the correlator, we transform it in a different form to make the terms better accessible for a physical interpretation.
We start by rewriting $C_k$ as
\begin{equation}
	C_k = A_k - 2 \left<\left<p_{\rm T}\right>_k\right> (N_k-1)\sum_{i=1}^{N_k} p_{{\rm T},i}^{(k)}+N_k(N_k-1)\left<\left<p_{\rm T}\right>_k\right>^2
\end{equation}    
with
\begin{align}
A_k &=\mathop{\sum_{i,j=1}^{N_k}}_{i\neq j} (p_{{\rm T},i}^{(k)}\quad p_{T,j}^{(k)})
= \left(\sum_{i=1}^{N_k} p_{{\rm T},i}^{(k)}\right)^2 - \sum_{i=1}^{N_k} \left( p_{{\rm T},i}^{(k)}\right)^2 \nonumber\\
&= \left<p_{\rm T}\right>_k^2N_k^2-\left<p_{\rm T}^2\right>_k N_k
\end{align}
Now we calculate $C_k/(N_k(N_k-1))$:
\begin{align}
\frac{A_k}{N_k(N_k-1)} 
&= \frac{\left<p_T\right>^2_k N_k - \left<p_T^2\right>_k}{N_k-1},
\end{align}
\begin{align}
- \frac{2 \left<\left<p_{\rm T}\right>_k\right> (N_k-1)\sum_{i=1}^{N_k} p_{T,i}^{(k)}}{N_k(N_k-1)} 
&=  - 2 \left<\left<p_{\rm T}\right>_k\right>\left<p_{\rm T}\right>_k,
\end{align}
\begin{align}
\frac{N_k(N_k-1)\left<\left<p_{\rm T}\right>\right>^2}{N_k(N_k-1)}
= \left<\left<p_{\rm T}\right>_k\right>^2 .
\end{align}

\noindent
Next, we introduce the variance $\sigma^2_{p_{\rm T},k} = \left<p_{\rm T}^2\right>_k - \left<p_{\rm T}\right>^2_k$ of the transverse momentum distribution of event $k$. 

\noindent
We rewrite $A_k/ (N_k(N_k-1))$ as: 
\begin{align}
\frac{A_k}{N_k(N_k-1)} = \frac{\left<p_{\rm T}\right>^2_k N_k - \left<p_{\rm T}^2\right>_k}{N_k-1} = \left<p_{\rm T}\right>^2_k - \frac{\sigma^2_{p_{\rm T},k}}{N_k-1}
\end{align}
\noindent
Putting everything back into the definition:
\begin{align}
\left<\Delta p_{{\rm T},i}, \Delta p_{{\rm T},j}\right>
&= \frac{1}{N_{ev}} \sum_{k=1}^{N_{ev}} \frac{A_k}{N_k(N_k-1)} - \left<\left<p_{\rm T}\right>_k\right>^2 \nonumber\\
&= \frac{1}{N_{ev}} \sum_{k=1}^{N_{ev}} \left<p_T\right>_k^2 - \frac{1}{N_{ev}} \sum_{k=1}^{N_{ev}} \frac{\sigma^2_{p_T,k}}{N_k-1} - \left<\left<p_{\rm T}\right>_k\right>^2 \nonumber\\
&= \left<\left<p_{\rm T}\right>_k^2\right> -\left<\left<p_{\rm T}\right>_k\right>^2 - \left<\frac{\sigma^2_{p_{\rm T},k}}{N_k-1}\right>\nonumber\\
&={\rm Var}(\left<p_{\rm T}\right>_k) -\left<\frac{\sigma^2_{p_{\rm T},k}}{N_k-1}\right> .
\end{align}\label{eq:sum}
Where the averaging is over all $k$ events. 
The interpretation of these terms is now clear: The term ${\big<\left<p_{\rm T}\right>_k^2\big> - \big<\left<p_{\rm T}\right>_k\big>^2}$ is the variance of the mean transverse momentum of the events, while the term ${\big<{\sigma^2_{p_{\rm T},k}}/{(N_k-1)}\big>}$ is the averaged scaled variance of the transverse momentum distribution in the single events. It is immediately clear that both terms are positive but either term can be larger which means that the covariance can take either positive or negative values or vanish altogether. 
Without any correlations, the Var$(\left<p_{\rm T}\right>_k)$ of the mean transverse momenta is related to variance of the in-event $\sigma^2_{p_{{\rm T},k}}$ by Var$(\left<p_{\rm T}\right>_k)= \EX\left[\sigma^2_{p_{{\rm T},k}}/N_k\right]$ for events without any correlations. In this case the correlator simply vanishes for $N_k\rightarrow \infty$.

In the above we have assumed that particles are drawn independently from a single particle distribution function (PDF). In such a case, getting negative values for the covariance is possible due to the sampling error in determining $\left<\left<p_{\rm T}\right>_k\right>$ which will give a negative contribution which, however, scales with one over the total number of particles used in the analysis.

As speculated by STAR, a critical point then may lead to a substantial correlation of the system and may introduce large fluctuations of the variance from event to event and thus a non-monotonic behavior of the covariance. While this is a very desirable option, one needs to rule out alternative explanations before one can draw such a strong conclusion. 

In the following, we will show that the transition from a baryonic system at low energies to a mesonic system at high energies can also create such a behavior.

STAR uses the correlator for charged particles. To first approximation these are protons and charged pions. Both particle species have significantly different mean transverse momenta. The relative contributions of protons and pions to the total charged hadron multiplicity depends, of course, on the collision energy. In general, within a fixed mid-rapidity window, the proton number is going down and the pion number is going up as the collision energy increases. Thus, one turns from a purely proton system to a purely pion system with increasing energy. However, at some intermediate energy, the proton and pion numbers will be similar. Naively, one may think that if pions and protons are drawn independently, that the covariance would also vanish.

There is, however, another subtlety that can have a significant effect.

\subsection{Effect of two different particle sources}

In their analysis, STAR uses charged particles for the calculation of the covariance. These charged particles consist of different species, pions, protons, Kaons etc. Each of these species follows its own momentum distribution and the relative yields of these species shows a strong beam energy dependence.
Let us assume a simple model system that is composed out of two instead of one particle species, and further assume that these particles have different momentum distributions.

Assume that each event contains fixed particle numbers $N_M$ and $N_B$ with $N_{k}=N_M+N_B$,
with mesons and baryons sampled independently from two different single-particle transverse momentum distributions with means:

\begin{equation}
\mu_M=\langle p_{\rm T}\rangle_M,
\qquad
\mu_B=\langle p_{\rm T}\rangle_B.
\end{equation}

The global mean transverse momentum is

\begin{equation}
\langle\langle p_{\rm T}\rangle_k\rangle=\frac{N_M\mu_M+N_B\mu_B}{N_K} = \bar p_{\rm T}.
\end{equation}

Using eq.(\ref{eq:dptdpt}) the pair sum can be separated into meson-meson, baryon-baryon and mixed meson-baryon contributions:

\begin{align}
\left\langle \Delta p_{{\rm T},i}\Delta p_{{\rm T},j}\right\rangle &=
\frac{N_M(N_M-1)\left\langle (p_{\rm T}^M-\bar p_T)(p_{\rm T}^M-\bar p_{\rm T}) \right\rangle}{N_k(N_k-1)} \nonumber\\
&+ \frac{N_B(N_B-1)\left\langle (p_{\rm T}^B-\bar p_{\rm T})(p_{\rm T}^B-\bar p_{\rm T}) \right\rangle}{N_k(N_k-1)} \nonumber\\
&+\frac{2 N_M N_B \left\langle (p_{\rm T}^M-\bar p_{\rm T})(p_{\rm T}^B-\bar p_{\rm T}) \right\rangle}{N_k(N_k-1)}.
\end{align}

Since the particles are sampled independently we have,

\begin{equation}
\left\langle (p_{\rm T}^M-\bar p_{\rm T})(p_{\rm T}^M-\bar p_{\rm T})
\right\rangle = (\mu_M-\bar p_{\rm T})^2,
\end{equation}

\begin{equation}
\left\langle (p_{\rm T}^B-\bar p_{\rm T})(p_{\rm T}^B-\bar p_{\rm T})
\right\rangle = (\mu_B-\bar p_{\rm T})^2,
\end{equation}

and

\begin{equation}
\left\langle (p_{\rm T}^M-\bar p_{\rm T})(p_{\rm T}^B-\bar p_{\rm T})
\right\rangle = (\mu_M-\bar p_{\rm T})(\mu_B-\bar p_{\rm T}).
\end{equation}

Therefore,

\begin{align}
\left\langle \Delta p_{{\rm T},i}\Delta p_{{\rm T},j}\right\rangle
&=
\frac{N_M(N_M-1)(\mu_M-\bar p_{\rm T})^2}{N_k(N_k-1)}
\nonumber\\
&+
\frac{N_B(N_B-1)(\mu_B-\bar p_{\rm T})^2}{N_k(N_k-1)}
\nonumber\\
&+
\frac{2 N_M N_B(\mu_M-\bar p_T)(\mu_B-\bar p_{\rm T})}{N_k(N_k-1)}.
\end{align}

Using

\begin{equation}
N_M(\mu_M-\bar p_{\rm T}) + N_B(\mu_B-\bar p_{\rm T}) = 0,
\end{equation}

the numerator simplifies to

\begin{equation}
- \left(N_M(\mu_M-\bar p_{\rm T})^2 + N_B(\mu_B-\bar p_{\rm T})^2 \right).
\end{equation}

Thus,

\begin{equation}
\left\langle \Delta p_{{\rm T},i}\Delta p_{{\rm T},j}\right\rangle
= - \frac{
N_M(\mu_M-\bar p_{\rm T})^2 + N_B(\mu_B-\bar p_{\rm T})^2}
{N_k(N_k-1)}
\end{equation}

or equivalently

\begin{equation}
\left\langle \Delta p_{{\rm T},i}\Delta p_{{\rm T},j}\right\rangle = 
-\frac{N_M N_B}{N_{k}^{2}(N_k-1)}(\mu_M-\mu_B)^2
\end{equation}

This means that if we independently draw particles from two different contributions, the covariance will receive a negative contribution that is proportional to the square of the difference of the means of the two distributions. If the number of one of the two particle species becomes much larger than the other, we obtain again the result for a single particle distribution, i.e. the covariance vanishes.

As a result, the assumption of a monotonous energy dependence of the covariance is only true if one assumes only a single particle species because the variance (width) is connected to the mean transverse momentum of this particle species which varies only little with center of mass energy for central nucleus-nucleus reactions. 

\section{Comparison to data within a minimalistic model}
To make a quantitative connection to the data, we use a minimalistic approach to explore the main contributions to the correlator within a simple model. We will see that a dip in the scaled correlation function $\sqrt{\left<\Delta p_{{\rm T},i}\Delta p_{{\rm T},j}\right>}/\left<\left<p_{\rm T}\right>_k\right>$ similar to the one observed in the STAR data around $\sqrt{s_{NN}} \approx 4-10$ GeV can emerge easily. 

The model set-up to describe the Au+Au data is as follows: We assume that the charged particles $N_k$ as a function of collision energy are composed of charged pions $N_{\pi}$ and protons\footnote{Here, 'protons' summarizes protons plus anti-protons.} $N_p$. The specific values are obtained from a fit to UrQMD simulations within mid-rapidity $|y|<0.5$ and $0.2<p_{\rm T}<2$ GeV to match the experimental conditions. Next we assume a standard thermal transverse momentum distribution 

\begin{equation}
{dN/dp_{\rm T} \propto p_{\rm T} m_{\rm T} \exp(-m_{\rm T}/T_{\rm eff})},
\end{equation}
with ${m_{\rm T}^2=p_T^2+m^2}$. A different effective temperature is employed for pions and protons to mimic collective flow, the values are adjusted to match the $\left< p_{\rm T}\right>$ of pions and protons using different $T_{\rm eff}$. To keep the model simple, we neglect the small energy dependence of $T_{\rm eff}$.

\subsection{Momentum conservation}\label{mom}
Since the STAR data also shows a small positive and almost energy independent contribution to the correlator we need to introduce an effect that increases the first term of eq. \ref{eq:sum}. This effect can come for example from early local momentum conservation (and survive the re-scattering as shown in \cite{Gavin:2016nir}) or radial flow fluctuations \cite{Bozek:2017jog,Schenke:2020uqq}.

The effect of momentum conservation can be implemented in a simple way. If one assumes momentum is conserved pairwise we can obtain $\vec p_i = -\vec p_j$ for all pairs. Since we are looking at transverse momenta which have no negative sign, we can just duplicate each transverse momentum  
(i.e. ${p_{{\rm T},1},p_{{\rm T},1},p_{{\rm T},2},}$ ${p_{{\rm T},2}, ..., p_{{\rm T},m_k},p_{{\rm T},m_k}}$) 
in a given event.

Assuming $N_k=2m_k$ in every event is fixed and the momenta $p_{{\rm T},i}^{(k)}$ in event $k$ are independent, i.e. particles are drawn in pairs with equal $p_{\rm T}$, then the effective number of independent random draws is reduced by a factor of 2. The event mean $p_{\rm T}$ and the variance ${\rm Var}(p_{{\rm T},i}^{(k)})=\sigma^2_{p_{\rm T},k}$ is not changed because the single particle distribution is not changed. 
Assuming (to make the argument clear) that the single particle distribution is the same in all events, the variance of the event means however is changed: ${{\rm Var}(\left<p_{\mathrm{T},k}\right>) =\sigma^2_{p_{\rm T},k}/m_k = 2\sigma^2_{p_{\mathrm{T,}}k}/N_k}$ which means it is increased by a factor of 2.

Thus, momentum conservation will create a positive contribution to the covariance.

It is clear that in reality, $N_k$ and $\sigma^2_{p_{\rm T},k}$ fluctuate from event to event and might even be (anti-)correlated, making a quantitative analytical estimate impossible.

\begin{figure} [t!]
    \centering
    \includegraphics[width=\columnwidth]{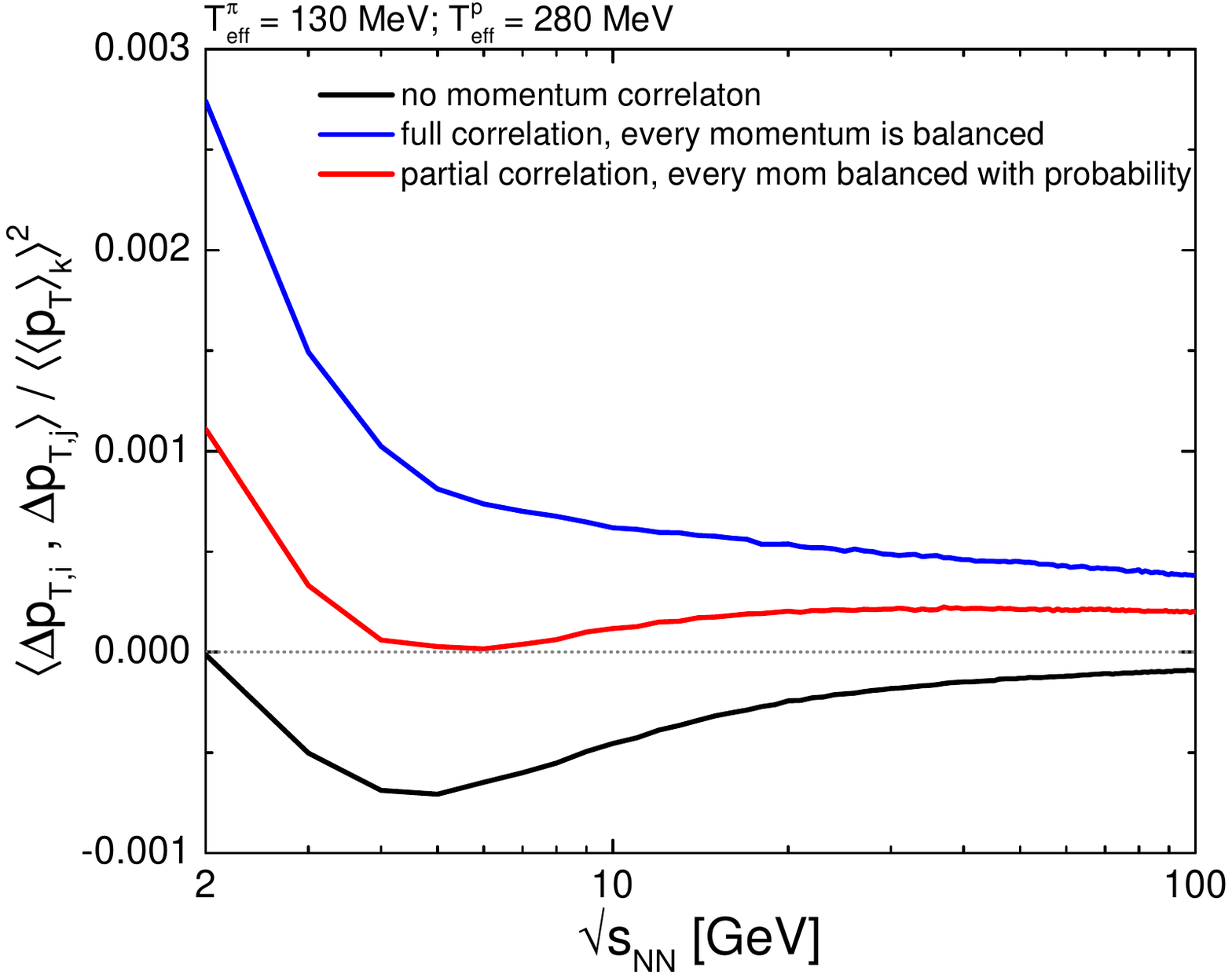}
    \caption{The scaled correlator \texorpdfstring{$\left<\Delta p_{{\rm T},i}, \Delta p_{{\rm T},j}\right>/\left<\left< p_{\rm T}\right>_k\right>^2$}{<Delta pT,i, Delta pT,j>/< <pT>_k >^2} (this is the square of the quantity shown by STAR to allow for negative values of the correlator) as a function of collision energy for \texorpdfstring{$T^p_{\rm eff}=0.280$}{Tp_eff=0.280} GeV, \texorpdfstring{$T^{\pi}_{\rm eff}=0.130$}{Tpi_eff=0.130} GeV without momentum conservation (black line), with full momentum conservation (blue line) and with a realistic momentum conservation (red line) probability as discussed in the text.}
    \label{fig:corr-uncorr-real}
\end{figure}

In our simplistic model we implement the momentum conservation by drawing $N_p/2$ transverse momenta from the thermal distribution for proton pairs. However, the balancing momentum may also be carried by a neutron (which is not taken into account in the charged particle correlator). The momentum matching then only takes place with a probability of $Z/A =79/197$ for a gold nucleus. In the case of an odd number of protons, one proton momentum stays unmatched. For the pions we draw $N_{\pi}/2$ transverse momenta from the thermal distribution for pion pairs. Again, since the balancing momentum may also be carried by a neutral pion (which is not taken into account in the charged particle correlator) the momentum matching only takes place with probability $2/3$. In the case of an odd number of pions, one pion momentum stays unmatched.

We therefore have three scenarios for the correlations from momentum conservation: 1. Without conservation (no correlation). 2. Full pairwise conservation (full correlation). 3. Partial pairwise correlation (weaker correlation). 

Of course in real events or transport simulations $\sigma^2_{p_{\rm T},k}$ can be correlated with $N_k$ and the conservation laws are only approximately fulfilled when looking only at charged particles in a given acceptance. Also detector effects like efficiency and centrality determination may influence the quantitative result. Such detailed analysis can only be done knowing the exact detector response and properties and are not the goal of this work.

\begin{figure} [t!]
    \centering
    \includegraphics[width=\columnwidth]{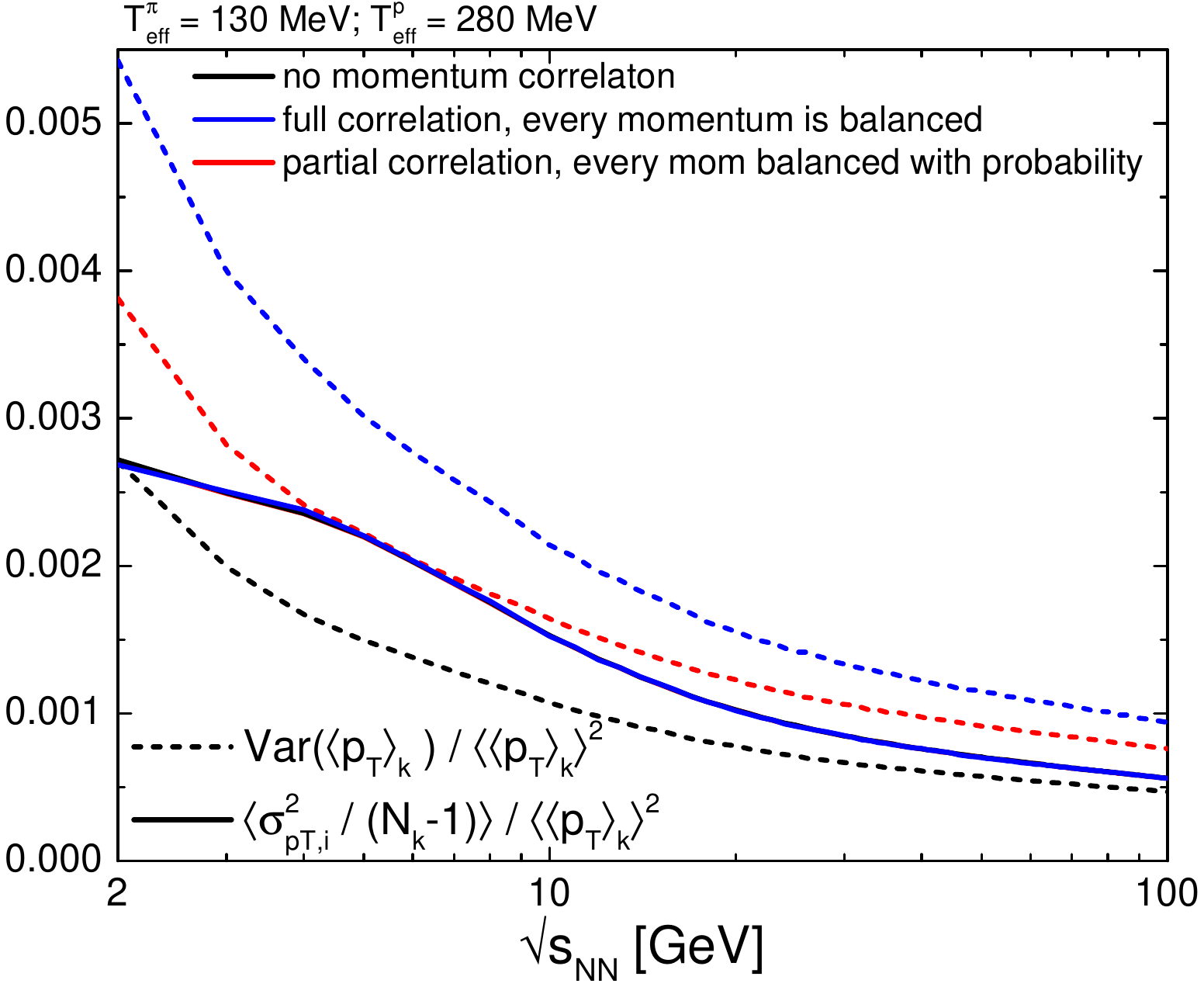}
    \caption{Energy dependence of the two components of the scaled correlator ${\left<\Delta p_{{\rm T},i}, \Delta p_{{\rm T},j}\right> = {\rm Var}(\left<p_{\rm T}\right>_k) - \left<\sigma^2_{p_{{\rm T},i}}/(N_k-1)\right>}$ for different scenarios of momentum conservation. 
    The black line denotes a scenario without any correlations and the blue line assumes full correlation. The red line corresponds to a more realistic scenario.
    The term ${Var(\left<p_{\rm T}\right>_k)/\left<\left< p_{\rm T}\right>_k\right>^2}$ (dashed lines) decreases with increasing collision energy, as expected for an increasing number of $N_{\rm ch}$. The term $\left<\sigma^2_{p_{{\rm T},i}}/(N_k-1)\right>/\left<\left< p_T\right>_k\right>^2$ (full lines) has a local maximum around a collision energy of $\sqrt{s_{NN}} \approx 4-10$ GeV but does not depend on the pair-wise correlation (all three lines are on top of each other).}
    \label{fig:cont_cons}
\end{figure}

\subsection{Results}

Let us start by exploring the effect of the correlation introduced by the momentum conservation. To this aim, we show in Fig. \ref{fig:corr-uncorr-real} the scaled covariance $\left<\Delta p_{{\rm T},i}, \Delta p_{{\rm T},j}\right>/\left<\left< p_{\rm T}\right>_k\right>^2$ as a function of collision energy for $T^p_{\rm eff}=0.280$ GeV, $T^{\pi}_{\rm eff}=0.130$ GeV without momentum conservation (black line), with full momentum conservation (blue line) and with a realistic momentum conservation probability (red line) as discussed above. One observes that the correlation introduced by the momentum conservation moves the covariance from negative values to positive values. Partial correlation reduced the value slightly compared to the fully correlated calculation. It is also clear that without any additional correlations the scaled covariance gives negative values in a beam energy range where the number of protons and pions becomes similar, as we expected. Already from these two trivial ingredients we have created a non-monotonous energy dependence and even negative values for the covariance are possible.

\begin{figure} [t!]
    \centering
    \includegraphics[width=\columnwidth]{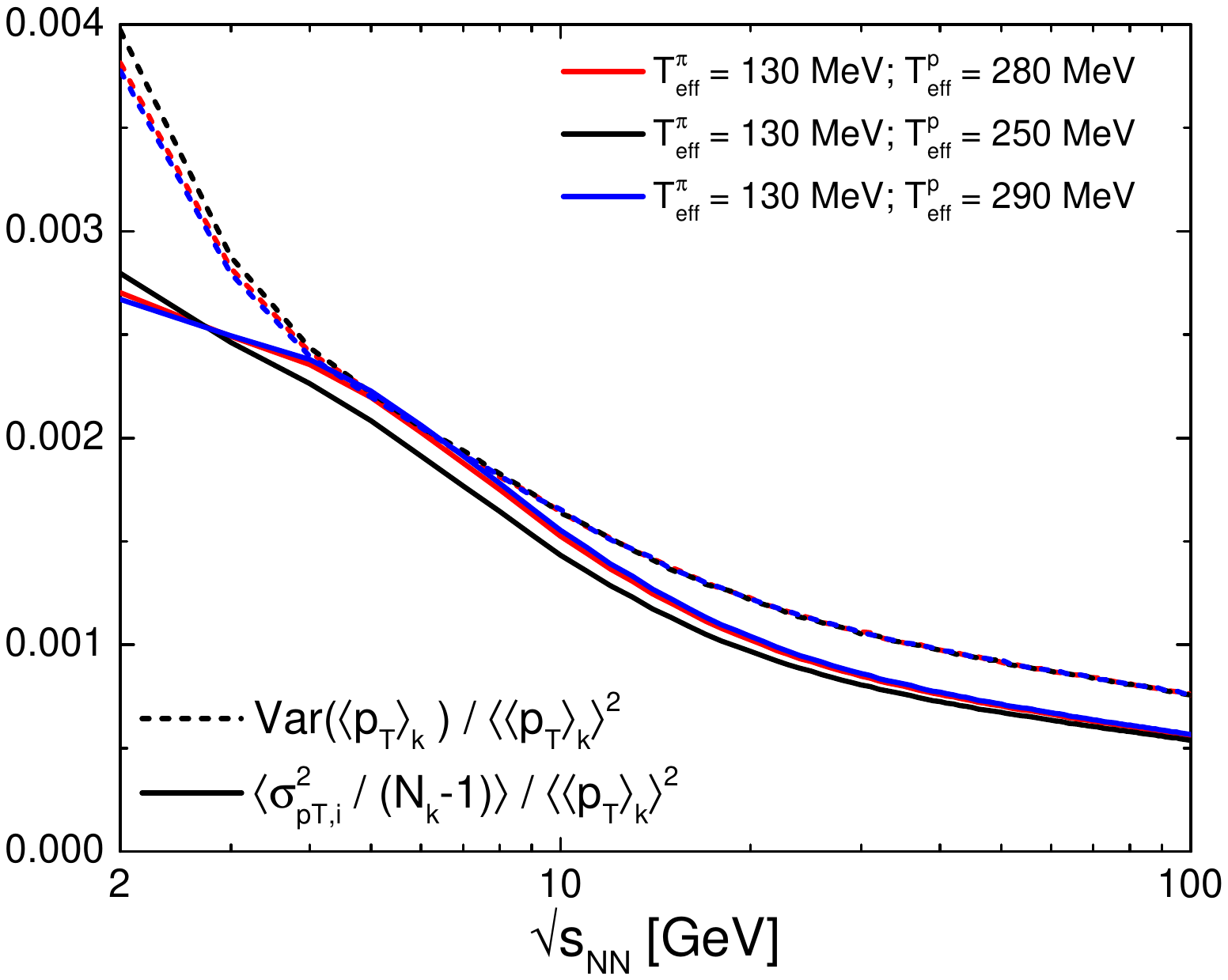}
    \caption{Energy dependence of the two components of the scaled correlator ${\left<\Delta p_{{\rm T},i}, \Delta p_{{\rm T},j}\right> = {\rm Var}(\left<p_T\right>_k) - \left<\sigma^2_{p_{T,i}}/(N_k-1)\right>}$ for different values for $T^p_{eff}=0.250, 0.280$ and $0.290$ GeV, keeping $T^{\pi}_{eff}=0.130$ GeV fixed. In all cases the partial conservation of momentum is assumed.    
    The term $Var(\left<p_T\right>_k)/\left<\left< p_T\right>_k\right>^2$  decreases with increasing collision energy but is almost independent of the choice of parameters. the term $\left<\sigma^2_{p_{T,i}}/(N_k-1)\right>/\left<\left< p_T\right>_k\right>^2$ shows a dependence on the the parameters of the momentum distributions which is strongest in the beam energy range of $\sqrt{s_{NN}} \approx 4-10$ GeV.}
    \label{fig:cont_dist}
\end{figure}

Next we turn in Fig. \ref{fig:cont_cons} to the collision energy dependence of the two components of the scaled covariance 
\begin{eqnarray}
\left<\Delta p_{{\rm T},i}, \Delta p_{{\rm T},j}\right>/\left<\left<p_{\rm T}\right>_k\right>^2 & = & {\rm Var}(\left<p_{\rm T}\right>_k)/\left<\left<p_{\rm T}\right>_k\right>^2 \nonumber \\
& - & \left<\sigma^2_{p_{{\rm T},i}}/(N_k-1)\right>/\left<\left<p_{\rm T}\right>_k\right>^2 \nonumber
\end{eqnarray}

In this figure, the three different scenarios for the correlation due to momentum conservation are compared. The case without conservation is shown as black lines, full conservation is shown as blue and the partial conservation is depicted as red lines. It is clear that the pairwise correlation affects only the first term that represents the variance of the mean transverse momenta. We also clearly see that a full pairwise sampling essentially doubles the value of the first term, as discussed above. The energy dependence can be understood by the simple scaling with the increasing $N_{\rm ch}$. The second term, is independent of this correlation, also scales with $N_{\rm ch}$, but shows a distinctive structure at an intermediate beam energy range due to the mixing of the pion and proton component, this means the energy region where $N_{\pi} \approx N_p$. 

To understand the origin and dependence of this structure better, Fig. \ref{fig:cont_dist} shows how the two terms  behave for different values for the parameters of the thermal distribution (i.e. we change the mean transverse momentum of the protons slightly): $T^p_{\rm eff}=0.250, 0.270, 0.280, 0.290$ GeV, while keeping $T^{\pi}_{\rm eff}=0.130$ GeV fixed. In this scenario we also assumed the partial conservation scenario as in the previous one. One observes that this time the first term is almost independent of the shapes of the two thermal probability distribution functions while the second term is modified slightly in the intermediate energy region.
This modification is enough so that the difference of these two components of the correlator becomes finite, close to zero or even turns negative. One should keep in mind that the asymptotic value of the root of the correlator divided by the mean of the mean $p_{\rm T}$ is on the one percent level.
Taken together, this explains why the full scaled correlator develops a minimum in this energy range, because the in-event $p_{\rm T}$ distribution is non-monotonic as a function of energy due to the transition from a proton to a pion system with a mixed distribution at intermediate energies. Our analysis also explains why transport models may provide strongly different results on this quantity, this is related to tiny (percent level) differences in the simulated variance of the in-event transverse momentum distribution that shows up as a drastic effect in the difference of the two components as discussed above.

\begin{figure} [t!]
    \centering
    \includegraphics[width=\columnwidth]{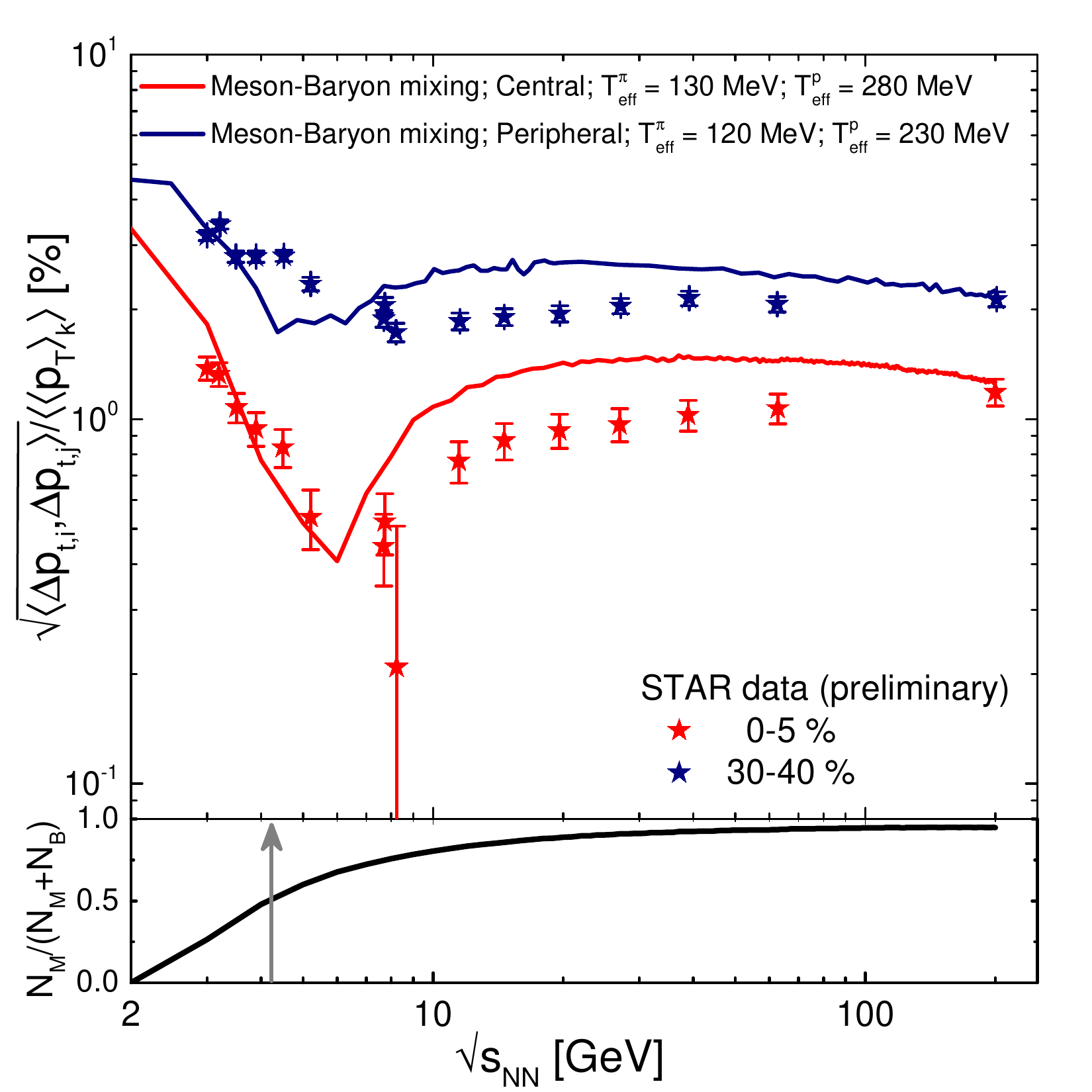}
    \caption{Energy dependence of $\sqrt{\left<\Delta p_{{\rm T},i}, \Delta p_{{\rm T},j}\right>}/\left<\left< p_{\rm T}\right>_k\right>$ for central collision (red line, using $T^p_{\rm eff}=0.280$ GeV, $T^{\pi}_{\rm eff}=0.130$ GeV) and for peripheral collisions (blue line, using $T^p_{\rm eff}=0.230$ GeV, $T^{\pi}_{\rm eff}=0.120$ GeV) in comparison to the STAR data \cite{STAR:2019dow,STAR:2026vjv}. Lower panel: Energy dependence of the meson fraction from UrQMD which is used as input for the simplistic model. The arrow indicates the beam energy where $50\%$ of all charged particles are mesons.}
    \label{fig:data-comparison}
\end{figure}

Finally, we compare to the experimental data measured by the STAR collaboration. In Fig. \ref{fig:data-comparison} we depict the calculations for $\sqrt{\left<\Delta p_{{\rm T},i}, \Delta p_{{\rm T},j}\right>}/\left<\left< p_T\right>_k\right>$ for central collision (using $T^p_{\rm eff}=0.280$ GeV, $T^{\pi}_{\rm eff}=0.130$ GeV) and for peripheral collisions (using $T^p_{\rm eff}=0.230$ GeV, $T^{\pi}_{\rm eff}=0.120$ GeV) in comparison to the STAR data. Mean Proton and pion yields for the different centralities and within the experimental acceptance are taken from the UrQMD model. One observes that the local minimum of the excitation function of the correlator is rather well reproduced in the minimalistic model. The minimum is correlated with the change from a bayonic to a mesonic system as can be seen in the lower panel of figure \ref{fig:data-comparison}.
Therefore we conclude that the change in the composition of the charged particles, accompanied by momentum conservation (or any other effect introducing a positive covariance) results in the structure observed by STAR and may not be related to the observation of the CEP of the phase transition line of QCD but is simply a result of the transition from a baryon to meson dominated system. 

\section{Conclusion}
It was shown that the observed minimum in the beam energy dependence of the charged particle transverse momentum covariance can be understood as an effect of mixing two different particle distributions in particular pions and protons. The minimum falls into the beam energy range where in heavy ion collisions, the transition between a baryon to meson dominated system occurs. For a simplistic two species system we provide the expected negative contribution to the scaled covariance from this mixing. To describe the data an additional small positive correlation is required. In our simulations this positive correlation comes from pairwise conservation of momentum; however, other effects like flow fluctuations are also known to give positive two particle correlations.

In a more realistic scenario the contributions to the covariance can become more complex and we should expect the following:

\begin{itemize}
\item Meson-baryon mixing: gives a negative contribution which is non-monotonic (as function of beam energy) and has a maximum at the beam energy where the number of mesons is the same as the baryons. This is likely the cause for the minimum at $\sqrt{s_{NN}}\approx 7.7$ GeV.
\item Pairwise momentum conservation: Gives a positive contribution due to the positive two particle correlation. The strength depends on how momentum is exactly balanced and the effect scales with $1/N_{ch}$ which makes it mostly a monotonic contribution. 
\item Fluctuations of $N_{ch}$, $N_{M}$ and $N_{M}$ which can be correlated. These are due to centrality fluctuations as well as random fluctuations of the number of particles in acceptance. In general, one would expect $N_M/N_B$ to be fluctuating not just randomly since meson production is coupled to baryons. This effect reduces the overall magnitude of the covariance, bringing it closer to zero. This also would lead to a mostly monotonous effect.
\item Finite statistics which leads to an error in the global mean of the transverse momentum leads to a systematic negative contribution which scales with 1 over the total number of charged particles summed over all events. If there are significant changes in the number of events considered as function of beam energy this effect can lead to a non-monotonic energy dependence. However, since the effect is very small it is unlikely to be the reason for the observed 'dip'.
\item Flow and flow fluctuations should introduce a positive correlation between particles in a given event and thus lead to a positive contribution to the covariance. This effect may depend on the beam energy and can be very model dependent which would require detailed modeling. However, it is unlikely that this effect would lead to any non-monotonicity in the $p_T$ fluctuations.
\end{itemize}

To summarize, we found that the non-monotonic beam energy dependence can be explained by trivial effects that are expected to occur in heavy ion reactions even without any critical point. The main effect which leads to the observed minimum at a specific beam energy range is the mixing of baryon and meson momentum distributions. Since the charged particle transverse momentum covariance turns out to be sensitive to a fine tuning of many aspects of the underlying event ensemble it is not a suitable observable to find the QCD critical endpoint.

\section*{Acknowledgments}
The authors thank Volker Koch, Rutik Manikandhan and Rene Bellwied for inspiring discussions. 
T.R. gratefully acknowledges support from The Branco Weiss Fellowship - Society in Science, administered by the ETH Z\"urich.

\bibliographystyle{elsarticle-num} 
\bibliography{main}

\end{document}